\documentclass[conference]{IEEEtran} 

\IEEEoverridecommandlockouts
\usepackage{cite}
\usepackage{amsmath,amssymb,amsfonts}
\usepackage{algorithm}
\usepackage{algorithmic}
\usepackage{graphicx}
\usepackage{textcomp}
\usepackage{xcolor}
\usepackage{bm}
\usepackage{subfigure}
\def\BibTeX{{\rm B\kern-.05em{\sc i\kern-.025em b}\kern-.08em
    T\kern-.1667em\lower.7ex\hbox{E}\kern-.125emX}}

\definecolor{red}{rgb}{1,0,0}
\usepackage{color,soul}
\definecolor{lightblue}{rgb}{.90,.95,1}
\sethlcolor{yellow} 

\begin{document}

\title{Decoder Error Propagation Mitigation for Spatially Coupled LDPC Codes}



\author{%
   \IEEEauthorblockN{Min Zhu\IEEEauthorrefmark{1},
                     David G. M. Mitchell\IEEEauthorrefmark{2},
                     Michael Lentmaier\IEEEauthorrefmark{3},
                     and Daniel J. Costello, Jr.\IEEEauthorrefmark{4}
                    }
   \IEEEauthorblockA{\IEEEauthorrefmark{1}%
                     \small{State Key Laboratory of ISN, Xidian University, Xi'an, P. R. China,
                     zhunanzhumin@gmail.com}}
   \IEEEauthorblockA{\IEEEauthorrefmark{2}%
                     \small{Klipsch School of Electrical and Computer Engineering, New Mexico State University, Las Cruces, NM, USA,
                     dgmm@nmsu.edu}}
   \IEEEauthorblockA{\IEEEauthorrefmark{3}%
                     \small{Department of Electrical and Information Technology, Lund University, Lund, Sweden,
                     michael.lentmaier@eit.lth.se}}
   \IEEEauthorblockA{\IEEEauthorrefmark{4}%
                     \small{Department of Electrical Engineering, University of Notre Dame, Notre Dame, IN, USA,
                     dcostel1@nd.edu}}
 }

\maketitle

\begin{abstract}
In this paper, we introduce two new methods of mitigating decoder error propagation for low-latency sliding window decoding (SWD) of spatially coupled low density parity check (SC-LDPC) codes. Building on the recently introduced idea of \emph{check node (CN) doping} of regular SC-LDPC codes, here we employ \emph{variable node (VN) doping} to fix (set to a known value) a subset of variable nodes in the coupling chain.  Both of these doping methods have the effect of allowing SWD to recover from error propagation, at a cost of a slight rate loss. Experimental results show that, similar to CN doping, VN doping improves performance by up to two orders of magnitude compared to undoped SC-LDPC codes in the typical signal-to-noise ratio operating range. Further, compared to CN doping, VN doping has the advantage of not requiring any changes to the decoding process.In addition, a log-likelihood-ratio based \emph{window extension algorithm} is proposed to reduce the effect of error propagation. Using this approach, we show that decoding latency can be reduced by up to a significant fraction without suffering any loss in performance.
\end{abstract}


\section{Introduction}
\emph{Spatially coupled low-density parity-check (SC-LDPC)} codes, a type of LDPC convolutional code \cite{LDPCCC1999}, have been shown to achieve \emph{threshold saturation}, i.e., the suboptimal \emph{belief propagation (BP)} iterative decoding threshold of SC-LDPC code ensembles over memoryless binary-input symmetric-output channels coincides with the \emph{maximum a posteriori probability (MAP)} threshold of their underlying LDPC block code (LDPC-BC) ensembles \cite{Lentmaier2010TIT,Kudekar2011TIT,Kudekar2013TIT,Kumar2012proof}. Further, \emph{regular} SC-LDPC code ensembles not only have capacity approaching iterative decoding thresholds, but they are asymptotically good, i.e., their minimum distance grows linearly with frame length \cite{David2015TIT}. Therefore, SC-LDPC codes combine the best features of both \emph{regular} and \emph{irregular} LDPC-BCs.

\begin{figure}
    \centering
    \includegraphics[width=0.4\textwidth]{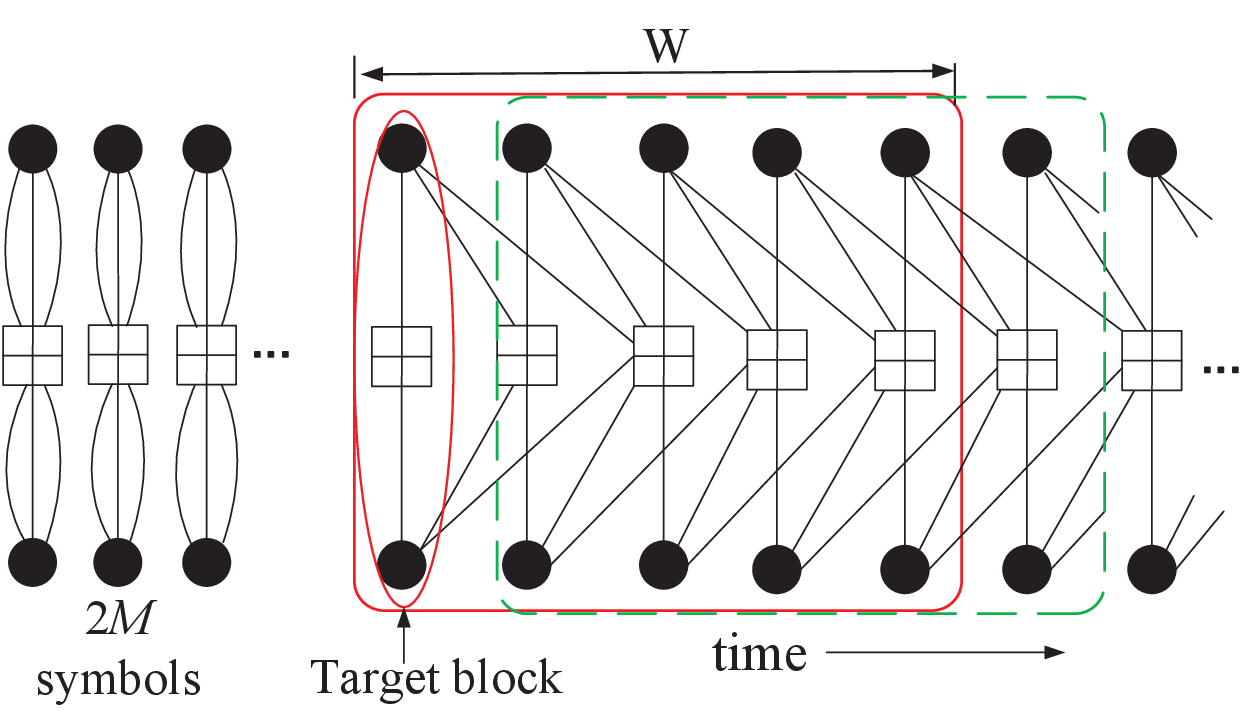}
\vspace{2mm}
\caption{A (3,6)-regular SC-LDPC code protograph obtained from an underlying LDPC-BC protograph with base matrix $B = \left[ {3,3} \right]$. The black circles represent variable nodes, and the ``plus'' squares represent check nodes. (a) A sequence of independent (uncoupled) protographs; (b) Spreading edges to the $m=2$ nearest neighbors.}\label{fig:Protograph}
\vspace{-5mm}
\end{figure}

SC-LDPC codes can be formed by applying a protograph-based construction technique \cite{David2015TIT}. In this paper we consider SC-LDPC codes constructed by coupling together a sequence of $L$ disjoint $(J,K)$-regular LDPC-BC protographs into a single coupled chain, where infinite $L$ results in an \emph{unterminated} coupled chain and finite $L$ results in a \emph{terminated} coupled chain. Without loss of generality, we consider an example of constructing (3,6)-regular SC-LDPC codes. We begin with an independent (uncoupled) sequence of (3,6)-regular LDPC-BC protographs with \emph{base matrix} ${\bf{B}} = \left[ {3,3} \right]$. Fig. \ref{fig:Protograph} shows the resulting unterminated (3,6)-regular SC-LDPC code chain obtained by applying the \emph{edge-spreading} technique of \cite{David2015TIT} to the uncoupled protographs. The edge spreading is defined by a set of \emph{component base matrices} ${{\bf{B}}_0} = {{\bf{B}}_1} = {{\bf{B}}_2} = \left[ {1~~1} \right]$ that must satisfy ${\bf{B}} = {{\bf{B}}_0} + {{\bf{B}}_1} + {{\bf{B}}_2}$. In general, an arbitrary edge spreading must satisfy ${\bf{B}} = \sum\nolimits_{i = 0}^m {{{\bf{B}}_i}}$,
where $m$ is referred to as the \emph{coupling width}.
Applying the lifting factor $M$ to the SC-LDPC protograph of Fig. \ref{fig:Protograph} results in an unterminated ensemble of (3,6)-regular SC-LDPC codes in which each time unit represents a \emph{block} of $2M$ coded bits (variable nodes).

To reduce the decoding latency and memory, \emph{sliding window decoding (SWD)} was proposed for SC-LDPC codes in \cite{Iyengar2012TIT}, where a standard BP flooding schedule is applied to all the nodes in the window. For example, in Fig. \ref{fig:Protograph}, the rectangular box represents a decoding window of \emph{size} $W$ (blocks). To decode, a BP flooding schedule is applied to all the nodes in the window for some fixed number of iterations, or until some \emph{stopping criterion} is met, the \emph{target block} of $2M$ symbols in the first window position is decoded according to the signs of their \emph{log-likelihood ratios} (LLRs), and the window shifts one time unit (block) to the right (see Fig. \ref{fig:Protograph}). Decoding continues in the same fashion until the entire chain is decoded, where the decoding latency in bits is given by $2MW$.

In order to reduce decoding latency and memory, the window size $W$ should be chosen as small as possible. In \cite{Huang2015TCOM} the authors experimentally showed that near optimal performance can be maintained at higher \emph{signal-to-noise ratios} (SNRs) as long as $W \ge 6\eta $, where $\eta=m+1$ is \emph{decoding constraint length}. When low latency operation is desired, however, at typical operating SNRs, smaller values of $W$ can sometimes result in infrequent but severe \emph{decoder error propagation}. Error propagation is triggered when, after a block decoding error occurs, the decoding of the subsequent block is also affected, which in turn can cause a continuous string of block errors, resulting in an unacceptable loss in performance. This is particularly damaging for very long code chains or for streaming applications.
Klaiber et al. in \cite{Klaiber2018ISTC} proposed to adapt the number of decoder iterations and/or shift the window position in order to combat decoder error propagation for SC-LDPC codes. For a related class of spatially coupled codes, viz. \emph{braided convolutional codes (BCCs)}, with SWD \cite{Zhu2017TCOM}, a \emph{window extension algorithm}, a \emph{synchronization} mechanism, and a \emph{retransmission} strategy were all used to mitigate error propagation \cite{Zhu2018ISIT}. More recently, Zhu et al. proposed a \emph{check node (CN) doped} SC-LDPC code design \cite{Zhu2020ISIT} to limit error propagation.  A disadvantage of these approaches is that they all require some modification of the decoding process.

The CN doped code design was motivated by the fact that the boundaries of a coupled chain have lower degree CNs, which has the effect of propagating more reliable information throughout the chain during iterative decoding. Inserting occasional lower degree CNs in a code chain, i.e., CN doping, has the same effect, thus allowing the decoder to recover from error propagation, although the shape of the decoding window must be altered at the doping points.  

In a similar vein, known (or fixed) variable nodes in a coupled chain can also aid the iterative decoding process.  This motivates us to propose a new class of \emph{variable node (VN) doped} SC-LDPC codes in this paper, which operate by inserting occasional fixed (known) VNs in a code chain, thus allowing the decoder to recover from error propagation, with the added advantage of leaving the shape of the decoding window unchanged. The CN doping and VN doping code designs introduce occasional irregularities in the coupled chain, which results in some rate loss. However, since code doping is primarily useful for long or unterminated chains, the rate loss associated with doping is very slight.\footnote{Since termination itself serves to truncate error propagation, the effect on short chains is minimal.}
We present numerical results showing that, similar to CN doping, VN doping improves performance by up to two orders of magnitude compared to undoped SC-LDPC codes in the typical SNR operating range. 

We also adapt the \emph{log-likelihood ratio (LLR)} based window extension algorithm, first proposed in \cite{Zhu2017TCOM} for BCCs, to combat error propagation in SWD of SC-LDPC codes. This approach, which requires the size of the decoding window to occasionally be extended, but involves no rate loss, is shown to be capable of reducing the decoding latency by a significant fraction without suffering any loss in performance.

\section{Decoder Error Propagation}
\begin{figure}[htbp]
   \centering
   \includegraphics[width=0.3\textwidth]{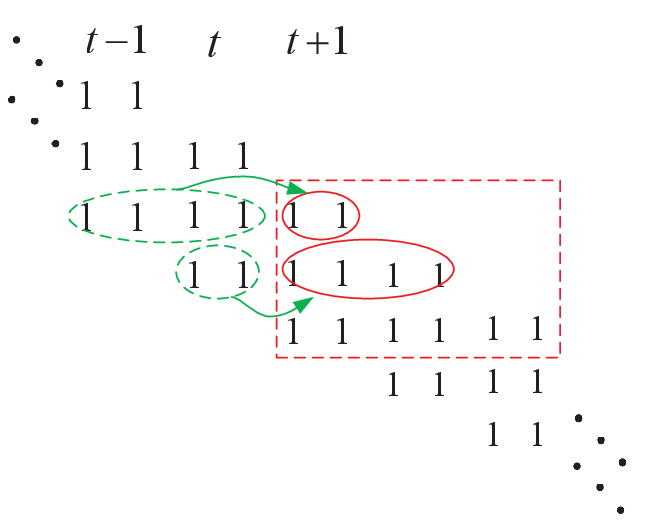}
   \caption{The final variable node LLRs at times $t-m+1$ to $t$ are used to update the check nodes in the window during the decoding of the target symbols at time $t+1$.}
   \label{fig:LLR_helpInWin}
 \end{figure}

During SWD of SC-LDPC codes, when a block of target symbols at time $t$ is decoded, the window shifts to include the most recent block of received symbols at time $t + W$, and decoding commences on the block of target symbols at time $t + 1$.  During the decoding of the time $t + 1$ block, for a coupling width of $m$, the final LLRs of the $m$ past decoded blocks, from time $t-m+1$ to time $t$, remain involved in the decoding process, although these LLRs are no longer updated, as illustrated in Fig. \ref{fig:LLR_helpInWin}, which depicts the base parity-check matrix of a (3, 6)-regular SC-LDPC code with $W=3$ and $m=2$.

Under normal operation, decoding proceeds with correctly decoded blocks until such time as a block of target symbols contains one or more LLRs with incorrect signs when the window shifts, thus resulting in a \emph{block decoding error}. Typically, if only a few symbols have incorrect LLRs and most of the correct LLRs have large magnitudes, the LLRs of the incorrectly decoded block will have only a small effect on the decoding of the next block, and the decoder will recover and continue to correctly decode subsequent blocks, assuming most of the symbols in the window have large and correct LLRs. This type of operation results in randomly distributed error blocks.

However, if an error block contains many incorrect LLRs, particularly if they have large magnitudes and a significant number of the LLRs associated with the correct symbols are small, those ``bad'' LLRs may negatively affect the decoding of the next block of target symbols, causing a block error that would not have occurred under normal operating conditions. This in turn can trigger additional block errors, resulting in an \emph{error propagation} effect, i.e., a continuous sequence of incorrectly decoded blocks.

In an application where information is transmitted in \emph{frames} of a small fixed length $L$ (in time units), with graph termination (reduced check node degrees for $m$ time units) following the last block of transmitted variable nodes, any error propagation will be limited and decoding will start fresh with the next frame. However, if $L$ is large, a significant number of blocks could be affected by error propagation, thus severely degrading performance. In a streaming application, with no termination, the situation could be catastrophic, resulting in a \emph{block error rate (BLER)} that asymptotically tends to 1.

\begin{figure}
\centering
   \includegraphics[width= 0.4\textwidth]{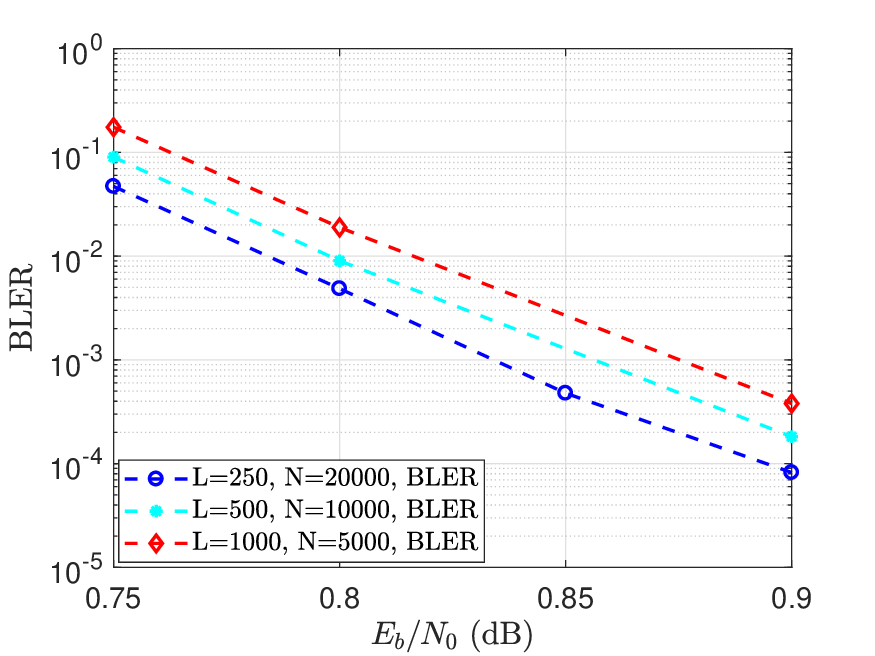}
   \caption{SWD BLER performance of a (3,6)-regular SC-LDPC code for three different combinations of frame length and number of frames simulated, all with the same total number of simulated blocks.}
   \label{fig:DiffNL}
   \vspace{-4mm}
\end{figure}

We now give an example illustrating the effect of this error propagation.
The simulated BLER performance of SWD of a (3,6)-regular SC-LDPC code based on the coupled protograph in Fig. \ref{fig:Protograph} is shown in Fig. \ref{fig:DiffNL}, where $W=18$ and $M=2000$. The figure represents the simulation of a total of $LN=5 \times {10}^{6}$ blocks (or $2MLN = 2 \times {10^{10}}$ bits), where $L$ is the frame length and $N$ is the number of frames simulated, for three different combinations of $L$ and $N$.
From the figure, we observe that, with increasing $L$, the BLER performance becomes worse, even though there are relatively few error-propagation frames overall, thus confirming the above observation.\footnote{Fig. \ref{fig:DiffNL} represents only a narrow range of SNRs, below the threshold of the underlying LDPC-BC, where error propagation presents a significant problem. For larger values of $E_b/N_0$ and/or $W$, SWD typically recovers from an error burst.}

\section{Error Propagation Mitigation}
In this section, we briefly review CN doping and then describe the new VN doping and window extension error propagation mitigation methods in detail.

\subsection{A Brief Review of CN Doped SC-LDPC Codes}
In order to combat error propagation in SWD of SC-LDPC codes, a new CN doped SC-LDPC code design was proposed in \cite{Zhu2020ISIT}. The key idea of CN doped SC-LDPC codes lies in occasionally inserting additional check nodes into the protograph of a regular SC-LDPC code, which is referred to as check node doping. The resulting structured irregularity limits error propagation by emulating graph termination, as noted earlier. We now use (3,6)-regular SC-LDPC codes as an example to briefly review of the CN doping process and the corresponding decoding scheme.

\begin{figure*}
\centering
 \includegraphics[width= 0.8\textwidth]{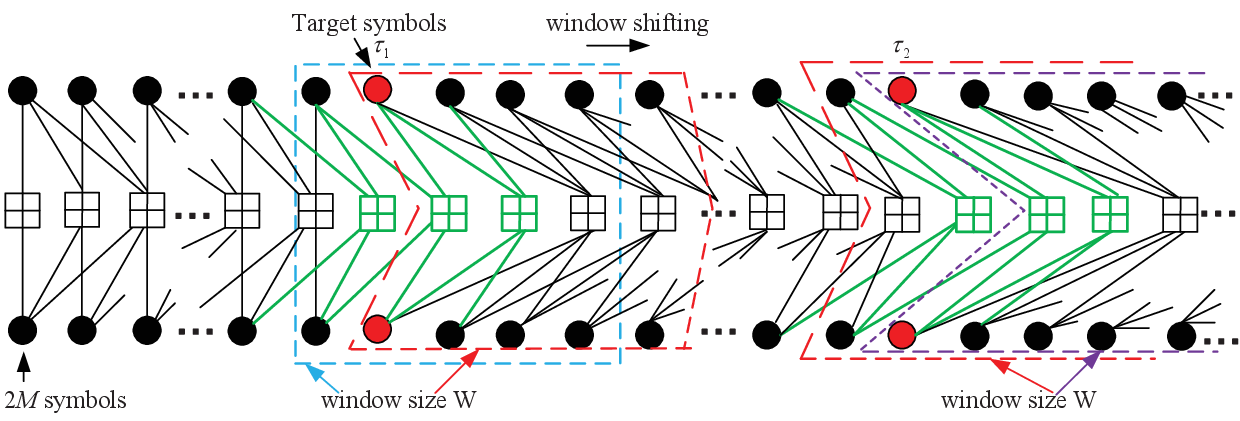}
\vspace{2mm}
\caption{The construction and SWD schedule for CN doped (3,6)-regular SC-LDPC codes.}
\label{fig:Matrix_review} 
\end{figure*}

Fig. \ref{fig:Matrix_review} shows the construction and SWD schedule of CN doped (3,6)-regular SC-LDPC codes. The red VNs at time $t={\tau}_1$ spread their three edges to the CNs at times $t={\tau}_1+1$, $t={\tau}_1+2$, and $t={\tau}_1+3$; the red VNs at time $t={\tau}_2$ spread their three edges to the CNs at times $t={\tau}_2+2$, $t={\tau}_2+3$, and $t={\tau}_2+4$, and so on.
To decode a CN doped (3,6)-regular SC-LDPC code, the window shifting schedule of SWD applied to the doped coupled chain is altered compared to standard SWD. When a doping point (red VN pair) becomes the target block, the window shifts by one VN time unit to include one new block of VNs, as before, but it shifts by two CN time units
to include two new blocks of CNs (and thus still including the same total number of CNs), as illustrated in Fig.  \ref{fig:Matrix_review}.

\subsection{VN Doped SC-LDPC Codes}
\begin{figure*}[htbp]
\centerline{\includegraphics[width= 0.8\textwidth]{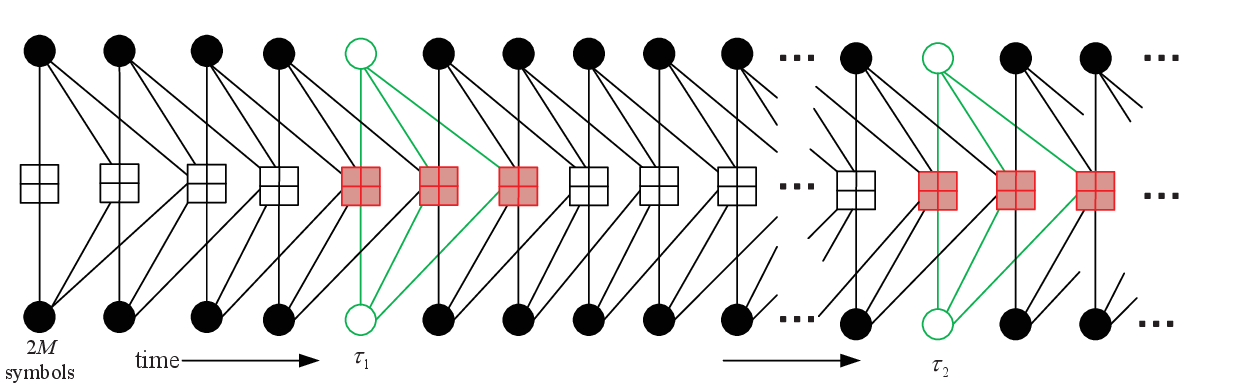}}
\caption{VN doping for a (3,6)-regular SC-LDPC code with occasional fixed variable nodes spaced throughout the coupled chain.}
\label{fig:VN_Doping}
\end{figure*}

As an alternative to introducing occasional reduced-degree check nodes in the coupled chain (called \emph{CN doping}) \cite{Zhu2020ISIT}, here the encoder fixes (set to ``0'') occasional variable nodes in the coupled chain, called \emph{VN doping}, as shown Fig. \ref{fig:VN_Doping}, where each time unit represents a \emph{block} of $2M$ coded symbols. The VNs at time $t={\tau}_1$ (the green empty circles) are doped by setting the $2M$ coded bits corresponding to these VNs to be ``0''. As a result, the CNs at times $t={\tau}_1, {\tau}_1+1, {\tau}_1+2$ (colored red and shaded) can be viewed as degree-4, rather than degree 6, CNs, thus emulating CN doping without actually altering the graph structure. Similarly, if the VNs at time $t={\tau}_2$ are doped, the CNs at times $t={\tau}_2, {\tau}_2+1, {\tau}_2+2$ (colored red and shaded) can be viewed as degree-4 CNs.

If $n_c$ and $n_v$ denote the total number of CNs and the total number of unknown VNs in VN doped SC-LDPC codes, respectively, and if there are $d$ doped positions, the design rate of VN doped SC-LDPC codes with frame length $L$ and $d$ doped VNs is
\begin{equation}
\begin{aligned}
{R_{L ,\rm{doped}}} = 1-\frac{{{n_c}}}{{{n_v}}} = 1-\left( {\frac{{L  + m}}{L-d}} \right)\left( {1-R} \right),
\label{eq:designR}
\end{aligned}
\end{equation}
where $R = 1 - {J}/{K}$ is the design rate of the uncoupled protograph \cite{David2015TIT}.
Compared to the design rate ${R_L } = 1 - \left( {\frac{{L  + m}}{L }} \right)\left( {1 - R} \right)$ of undoped SC-LDPC codes \cite{David2015TIT}, we see from \eqref{eq:designR} that the design rate of VN doped SC-LDPC codes is smaller, i.e., VN doping results in some \emph{rate loss}, similar to CN doping.

The decoding process is the same as for undoped codes, except that the doped code symbols and their positions are treated as known, i.e., during the decoding process we set the LLRs of the doped symbols to be a large constant negative value. These known bits have the effect of transmitting perfectly reliable information to their neighbour nodes, thus helping the decoder recover from error propagation. An important implementation advantage of VN doping over CN doping is that the shape of the decoding window remains unaltered.

\subsection{Window Extension Algorithm}
In this section, rather than altering the code design, we mitigate error propagation by occasionally extending the size of the decoding window.
By experimentally recording decoder behavior during error propagation, we find that the average LLR magnitudes of the blocks are typically near zero, a phenomenon also observed for BCCs in \cite{Zhu2018ISIT}.

To take advantage of this observation, we can allow the window size $W$ to change dynamically in SWD of SC-LDPC codes, from an initial size $W_{\rm{init}}$ to a maximum size $W_{\rm{max}}$, thereby causing additional decoding resources to be employed when a potential error propagation condition is detected. To formalize this process, we denote the decision LLRs of the $2M$ coded bits in the $i$th block of the current window after some fixed number $I$ of iterations by ${{\bm{\ell}} ^i} = \left( {\ell _0^i,\ell _1^i, \ldots ,\ell _{2M - 1}^i} \right)$, $i \in \left\{ {t,t + 1, \ldots ,t + W - 1} \right\}$. Then the average LLR magnitude of the $2M$ code bits in block $i$ after $I$ iterations is given by
\begin{equation}
\begin{aligned}
{{\bar {\ell} }^i} = \frac{1}{{2M}}\sum\limits_{j = 0}^{2M - 1} {\ell _j^i}.
\label{eq:AbsLLR}
\end{aligned}
\end{equation}
We also define the \emph{observation span} $\tau$ as the number of consecutive blocks in the decoding window over which the average LLRs magnitude is to be examined.

\begin{figure*}[htbp]
\centerline{\includegraphics[width= 0.8\textwidth]{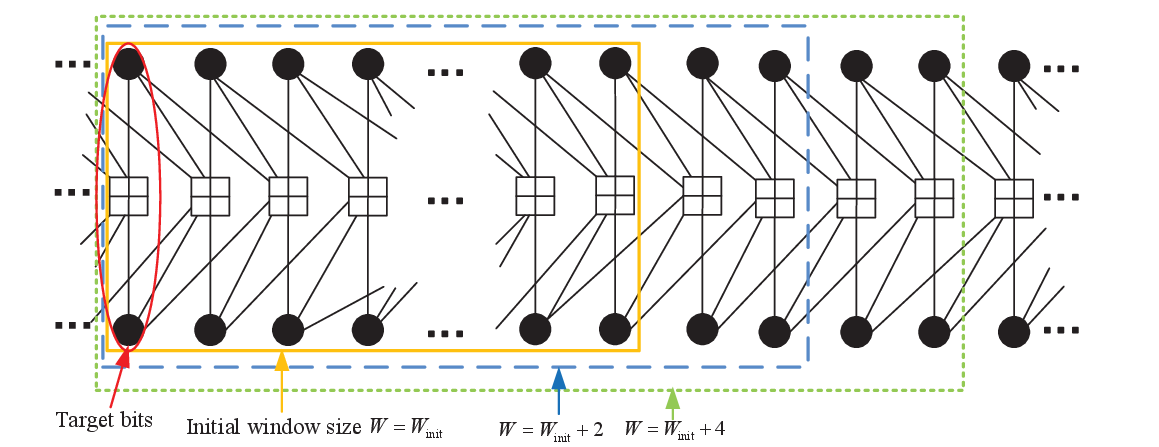}}
\caption{Sliding window decoder with the window extension algorithm for (3,6)-regular SC-LDPC codes.}
\label{fig:WinExtension}
\end{figure*}

Now assume the current decoding window, with $W=W_{\rm{init}}$, covers the blocks ($2M$ bits each) from time $t$ to $t+W-1$. After $I$ iterations, if any of the average LLR magnitudes of the first $\tau$ blocks in the current window, $1 \le \tau  \le W$, is lower than a predefined \emph{threshold} $\theta$, i.e., if
\begin{equation}
\begin{aligned}
{{\bar \ell }^i} < \theta, ~~~~~\mathrm{for~any}~i \in \left\{ {t,t + 1, \ldots ,t + \tau  - 1} \right\},
\label{eq:WinExt}
\end{aligned}
\end{equation}
then the target block is not decoded, the window size $W$ is increased by 2 time units\footnote{Increasing the window size by 2 time units was determined experimentally to be the best compromise between performance and complexity.}, and the decoding process restarts. If none of the first $\tau$ blocks is satisfies \eqref{eq:WinExt}, the target block is decoded and the window shifts by 1 time unit. If the current window size reaches $W_{\rm{max}}$, then the target block is decoded, the window shifts by 1 time unit, and the window size is reset to $W_{\rm{init}}$.\footnote{Note that, since the window size can vary, the decoding latency can be characterized by the average window size.}
Fig. \ref{fig:WinExtension} shows how the window extension scheme works during the SWD of (3,6)-regular SC-LDPC codes.
The method is described in detail in Algorithm 1.
\begin{algorithm}[!t]
\caption{Window Extension} \label{Alg:WinExt_Alg}
\begin{algorithmic}[1]
\STATE Assume that the block at time $t$ is the target block in a window decoder of size $W$ initialized with the channel LLRs of $w$ received blocks. Set $W=W_{\rm{init}}$ initially, and let $\tau$, $\theta$, $W_{\rm{init}}$, and $W_{\max}$ be parameters.

\STATE Every time $I_{\max}$ iterations are finished, calculate the absolute average LLR value of each block in the current window according to \eqref{eq:AbsLLR}.
  \IF{${{\bar \ell }^i} < {\theta }$ for any $i \in \left\{ {t,t+1, \ldots, t+\tau-1} \right\}$}
    \IF{$W < W_{\max}$}
     \STATE The decoder accepts two new blocks from the channel. The target block is still the block at time $t$, the new blocks are at times $t+W$ and $t+W+1$, and, the window size is set to $W=W+2$.
     \STATE For the old $W-2$ blocks in the window, all the LLRs are maintained.
     \STATE Restart decoding process, and go to step 2.
     \ELSE
     \STATE Go to step 12.
    \ENDIF
  \ELSE
  \STATE Decode the target block, set $W=W_{\rm{init}}$, and shift the window by one time unit.
  \STATE Continue decoding and go to step 2.
  \ENDIF
\end{algorithmic}
\end{algorithm}

\section{Numerical Results}
\begin{figure}[htbp]
\centerline{\includegraphics[width= 0.45\textwidth]{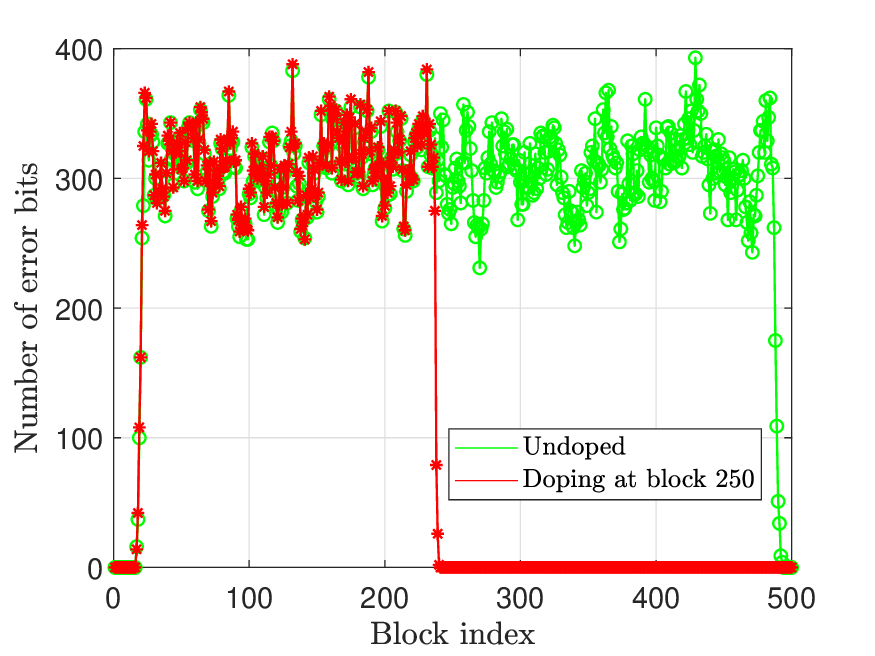}}
\caption{Bit error distribution per block.}
\label{fig:ErrDist}
\end{figure}

In order to illustrate the effectiveness of the proposed mitigation methods, the bit error distribution per block of a typical error-propagation frame in SWD of the VN doped and undoped (3,6)-regular SC-LDPC codes of Fig. \ref{fig:Protograph} sent over an AWGN channel with BPSK signaling is plotted in Fig. \ref{fig:ErrDist}, where $M=1000$, $L=500$, and $W=18$. The figure clearly shows that VN doping truncates the error propagation at the doping point in the center of the frame, whereas the errors continue to the end of frame in the undoped case. Similar behavior has been observed for CN doping (see \cite{Zhu2020ISIT}) and window extension.

\begin{figure}[htbp]
\centerline{\includegraphics[width= 0.45\textwidth]{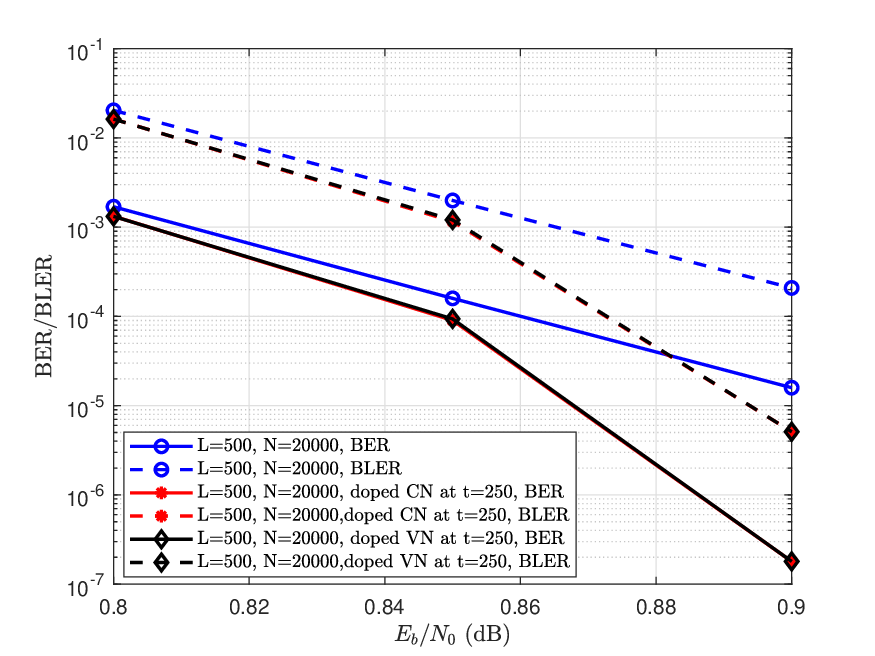}}
\caption{Performance comparison of CN doped, VN doped, and undoped SC-LDPC codes.}
\label{fig:Compare_VNCN}
\end{figure}

The \emph{bit error rate (BER)} and BLER performance of VN doping with rate $R=0.49699$, CN doping with rate $R=0.497$ \cite{Zhu2020ISIT}, and undoped with rate $R=0.499$ (3,6)-regular SC-LDPC codes of Fig. \ref{fig:Protograph} is shown in Fig. \ref{fig:Compare_VNCN}, where the termination length is $L=500$ and the window size is $W=18$. We observe that both VN doping and CN doping gain approximately two orders of magnitude in BER and one order of magnitude in BLER compared to the undoped code at those SNR operating points of interest (below the threshold of the underlying LDPC-BC). Also, the fact that the performance of both doping methods is essentially equivalent corroborates our earlier observation that VN doping emulates the CN doping process while not requiring any alteration to the shape of the decoding window.

\begin{figure}
\centering
\subfigure[$W=18$]{
        \includegraphics[width=0.45\textwidth]{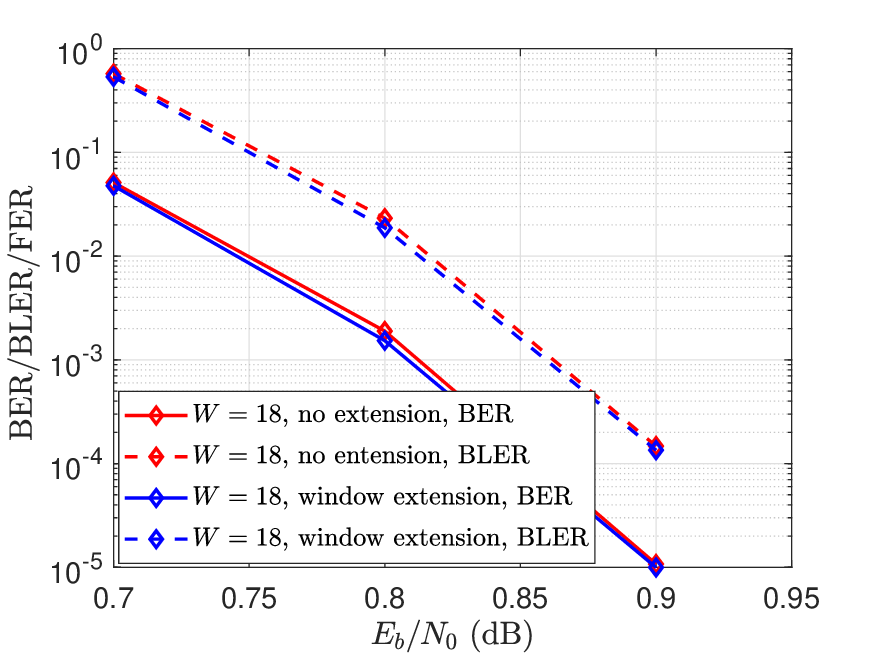}
        \label{fig:W18WinExtend}
    }
\subfigure[$W=9$]{
        \includegraphics[width=0.45\textwidth]{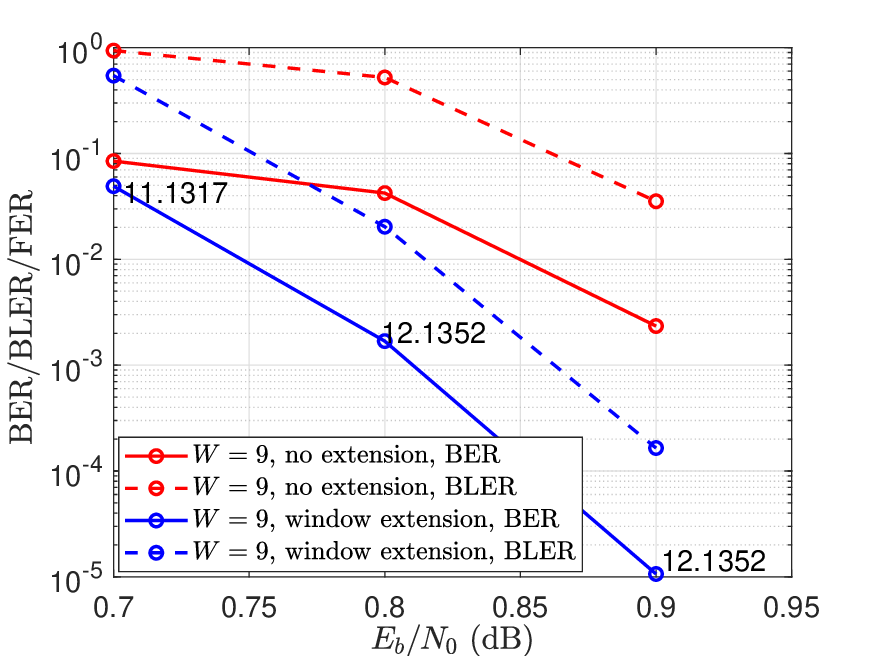}
        \label{fig:W9WinExtend}
    }
\vspace{2mm}
\caption{BER and BLER performance of window extension with window sizes $W=18$ and $W=9$.}
\label{fig:WindowExtend} 
\end{figure}

To illustrate the advantage of allowing a dynamic window size, Fig. \ref{fig:WindowExtend} shows the performance of the (3,6)-regular SC-LDPC codes of Fig. \ref{fig:Protograph} with and without window extension for $L=250$ and window sizes $W=18$ and $W=9$. From the figure, we see that, for $W=18$, window extension gives only a slight improvement compared to not using window extension. However, for $W=9$, window extension gains more than two orders of magnitude in both BER and BLER at SNR operating points of interest. From another point of view, by comparing Figs. \ref{fig:W18WinExtend} and \ref{fig:W9WinExtend}, we see that the performance with $W=9$ and window extension is roughly equivalent to that of $W=18$ with no window extension, while the average window size (decoding latency), indicated on the BER curve with window extension as shown in Fig. \ref{fig:W9WinExtend}, is reduced by about 1/3.

\section{Conclusions}
In this paper, we proposed two new methods of mitigating decoder error propagation in SWD of SC-LDPC codes: VN doping from the code design aspect and window extension from the decoder design aspect. The first, VN doping, takes advantage of occasional fixed (known) variable nodes in the protograph to allow SWD to recover from error propagation, without requiring any alteration to the shape of the decoding window (in contrast to the previously proposed CN doping). As a result, the BER and BLER performance of VN
doped SC-LDPC codes was shown to improve by up to two orders of magnitude at SNR operating points of interest. The second method, window extension, adapts the window size dynamically to bring more decoding resources to bear in error propagation conditions. Simulation results show that window extension with $W=9$ can reduce the decoding delay (latency) on average by about 1/3 compared to using $W=18$ without window extension, while maintaining the same BER and BLER performance.
%


\end{document}